# Serious errors impair an assessment of forest carbon projects: A rebuttal of West et al. (2023)

**Authors:** Edward T. A. Mitchard[1,2]*, Harry Carstairs[1], Riccardo Cosenza[3], Sassan S. Saatchi[4,5,6], Jason Funk[7], Paula Nieto Quintano[1], Thom Brade[1], Iain M. McNicol[1,2], Patrick Meir[2], Murray B. Collins[1], Eric Nowak[3,8]

**Affiliations:**

[1]Space Intelligence; 93 George Street, Edinburgh EH2, 3ES, UK

[2]School of GeoSciences, University of Edinburgh; Alexander Crum Brown Road, Edinburgh, EH9 3FF, UK.

[3]Center for Climate Finance and Sustainability, Università della Svizzera Italiana; Via Giuseppe Buffi 13, 6900 Lugano, Switzerland

[4]NASA Jet Propulsion Laboratory, California Institute of Technology, Pasadena, CA 91109, USA

[5]Dept. of Atmospheric Sciences, University of California, Los Angeles, CA 90095, USA

[6]CTrees, Pasadena, CA 91105, USA

[7]Conservation International; 2011 Crystal Drive, Suite 600 Arlington, VA 22202, USA

[8]Verified Carbon Standard (VCS) Advisory Group, Verra; Washington DC 20005, USA

*Corresponding author. Email: ed@space-intelligence.com

**Abstract:** Independent retrospective analyses of the effectiveness of reducing deforestation and forest degradation (REDD) projects are vital to ensure climate change benefits are being delivered. A recent study in *Science* by West et al. *(1)* appeared therefore to be a timely alert that the majority of projects operating in the 2010s failed to reduce deforestation rates. Unfortunately, their analysis suffered from major flaws in the choice of underlying data, resulting in poorly matched and unstable counterfactual scenarios. These were compounded by calculation errors, biasing the study against finding that projects significantly reduced deforestation. This flawed analysis of 24 projects unfairly condemned all 100+ REDD projects, and risks cutting off finance for protecting vulnerable tropical forests from destruction at a time when funding needs to grow rapidly.

**One-Sentence Summary:** The result of West et al. *(1)* that most REDD projects do not deliver carbon benefits is undermined by major flaws.

**Main Text:**

West et al. *(1)* estimate the effectiveness of 24 REDD projects from six countries, representing around a quarter of projects registered under the Verified Carbon Standard (VCS). The core of their method is to match each REDD project area (PA) to synthetic control (SC) sites: circular proxies of the same size as the PA that are assumed to undergo the same trend in deforestation as the PA would have, if the REDD project had not been implemented.

The area deforested in the PAs and SCs were estimated using a global scale forest loss dataset *(2)*, and West et al. attempted to quantify the carbon credits successfully delivered according to the difference between deforestation in the PAs and the SCs. They concluded that the majority of projects did not generate any carbon benefits, and of those that did, most issued more carbon credits than were justified. However, this analysis did not provide an accurate estimate of carbon benefits of these projects for three reasons:

> 1. The satellite-derived data *(2)* they use in all analyses has errors and biases that its creators state make it inappropriate for this type of comparison. These led to effective projects being incorrectly labeled as ineffective; and a potential underestimate of avoided deforestation of 90% or more.

> 2. Their project-level SC and country-level Generalized Synthetic Control (GSC) methods failed to select appropriate donor areas, with the SC sets failing common tests in 24 of 30 cases. The resulting models are also highly unstable, with fine-tuning of the method causing wildly different results. This is likely due to omitting key variables, such as 'distance to roads'.

> 3. West et al. made two different calculation errors, that mean the proportion of credits they report that deliver real carbon benefits should be increased by 62%.

West et al. further did not attempt to calculate the uncertainty on their estimates of carbon credits generated, nor did they adopt a formal hypothesis testing framework. These weaknesses have led to their results appearing more precise than they are, and may have led to a widespread misinterpretation that REDD projects are worthless *(3)*. This is compounded by extending their conclusions beyond their dataset: they generalize about REDD success on the basis of a non-random sample of 24 projects from a potential dataset of 103 registered Verra REDD projects *(4)*, itself a subset of 377 known REDD projects *(5)*.

Details on all these points follow below.

**Misuse of Global Forest Change Data**

West et al. based their analysis on the 'Global Forest Change' (GFC) dataset, a product designed for use as a "stratifier in targeting forest extent and/or change by a probability sample" *(2)*, e.g. as used by Tyukavina et al. *(6)*. The website where the data are accessed explicitly states that "definitive area estimation should not be made using pixel counts from the forest loss layers", and that the data does not meet standards for areal land-cover change estimation set by the Intergovernmental Panel on Climate Change *(7)*. However, West et al. did not adopt a point-based sample approach, instead using the data directly. This inevitably adds great uncertainty to their analysis, and due to the nature of their comparisons (between REDD PAs and either control sites or REDD PAs before their start date) reduces the chances that a successful project will be identified as such, or its benefits fully quantified.

The GFC data should not be used in this way because they contain false and missed detections that combine to reduce the apparent success of all projects. These errors can be large: for example, a study using field plots and Synthetic Aperture Radar data covering drier countries in sub-Saharan Africa, including all of Tanzania and Zambia, and the southern portion of DRC (3 of the 6 countries covered by West et al.) found the GFC data overestimated forest cover by a factor of two, but then underestimated forest loss rates by a factor of five, with these overall rates of error varying greatly by region and starting canopy cover *(8)*. We calculate that in this region this error rate would lead to an 89% understatement of project carbon benefits, meaning a project that stopped all deforestation (100% successful) would appear to have only stopped 11% if the GFC data were used to do the assessment. The African regional analysis in West et al. found an overall effectiveness of 16% (414 ha year$^{-1}$ or 423 ha year$^{-1}$ effect size, depending on method, compared to 2700 ha year$^{-1}$ baseline); the error rate in the GFC data used suggests this could really represent an effectiveness of well over 100%.

We calculate this based on assuming the detection rate ($R_d$ = proportion of true deforestation detected) and false alarm rate ($R_f$ = proportion of stable forest marked as deforestation) were the same in the PA and the SC, then the area of avoided deforestation would be underestimated by a factor ($R_d - R_f$). To understand this consider two sites with the same area $A$, of which $D_i$ is deforested: the predicted area of deforestation in site $i$ will be:

$$i = R_f(A - D_i) + R_d D_i \qquad \text{(Equation 1)}$$

The true difference in the areas of deforestation ($D_{Dif:1,2}$) predicted between sites 1 and 2 would therefore equal:

$$D_{Dif:1,2} = (D_1 - D_2)(R_d - R_f). \qquad \text{(Equation 2)}$$

For study *(8)* $R_d = 0.19$ and $R_f = 0.08$, leading to the 89% underestimate of effectiveness. But even using the estimates for wet tropical regions by the providers of the GFC data *(2)*, $R_d = 0.83$ and $R_f = 0.002$, leading to underestimation of real avoided deforestation by 17%. West et al could have corrected for this best case scenario by increasing their observed effectiveness in tables and figures by 17%, and recognising that the impact could be bigger in many areas.

These errors are further compounded because the satellites used and methodology applied in the GFC data have changed significantly during the time period analyzed by West et al. In particular a change in method and the use of Landsat 8 caused a big increase in detections post 2012, with particularly increased sensitivity to degradation and small-scale deforestation *(7, 9)*. This will have a spatially heterogeneous impact that may only become apparent while projects are in operation, but not during the training period of selecting Synthetic Controls. For example two sites, one with small-scale degradation and the other with no disturbance, might appear identical in GFC from 2000-2012, but diverge from 2013 onwards when the small-scale degradation starts being detected. This step change in sensitivity could significantly bias all the analyses in West et al.

**Invalid Implementation of Synthetic Control Method**

Leaving aside the quality of the underlying data, though it is itself sufficient to invalidate the results, we conducted an empirical evaluation of the synthetic control method (SCM) employed by West et al. SCMs were designed as a tool to understand the effects of state policy interventions *(10)*, and in a conservation context it is vital that steps are taken to account for domain specific complexity, select appropriate covariates, and test that valid controls have been obtained *(11)*. Here, we show that the approach taken by West et al. leads to inappropriate matching, ignores key drivers of deforestation, and that in most cases (24 of 30) the SCs set selected for PAs fail to pass standard validation tests.

*Synthetic Controls Poorly Constrained*

REDD baselines are created using reference regions that are located near to the project area, so that local laws, customs, forest characteristics, and pressures on forests are similar (though see Note 1). In contrast, West et al. draw potential donor areas from randomly generated circles over entire countries, in many cases leading to SCs that are incomparable. The covariates they use to generate SCs from the potential donors also lack critical variables: they do not include 'distance to roads' - identified by Busch & Ferretti-Gallon *(12)* as the single most important driver of deforestation risk. In fact, they only include two variables from the top five identified by Bush & Ferretti-Gallon, with 'Population' and 'Wealth' also omitted.

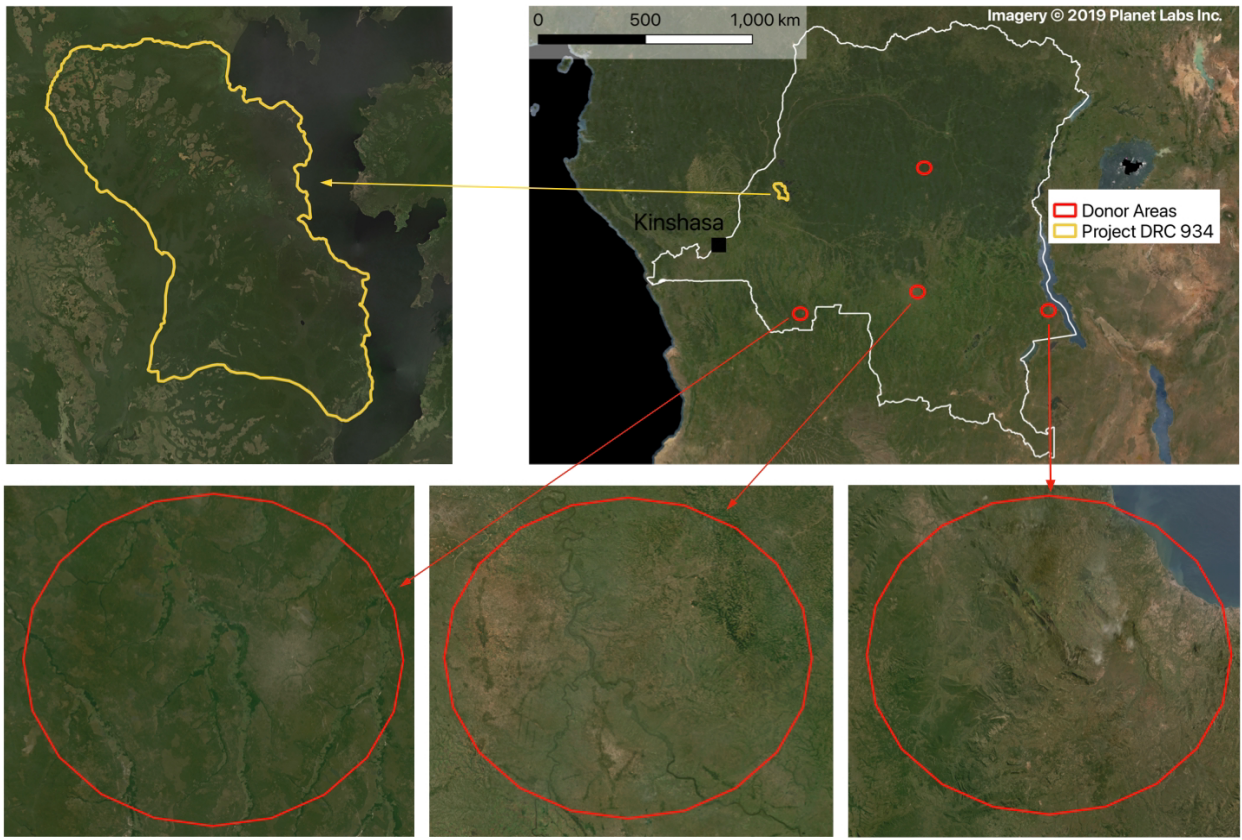

**Figure 1: Project area (PA) DRC 934 and the four donor areas used by West et al. *(1)* to generate its synthetic control (SC).** The white outline in the overview panel shows the border of the Democratic Republic of the Congo (DRC), while the zooms show the Project Area (in dense rainforest with small patches of savanna), and in the second row three of the four selected Synthetic Controls, where savanna dominates. The satellite basemap is from December 2019, with data provided by Planet and accessed through Norway's International Climate and Forest Initiative (NICFI).

The decision to include donor areas from entire countries leads to some poor comparisons. For example, two PAs in Peru and all Colombian PAs are compared to sites on the opposite side of the Andes (West et al. Fig. S4). Biomes, access to markets, diseases, and crops grown, are all vastly different on either side of the Andes in Peru and Colombia, all of which will influence deforestation dynamics.

Elsewhere, West et al. make comparisons across huge distances, not least in the DRC – a country with a similar land area to western Europe, and corresponding huge variance in modes and types of agricultural production, transport networks, access to markets, laws and culture. A simple comparison to satellite data shows that only 1 of the 4 selected donor areas for site DRC 934 is in a closed canopy forest like the PA, with the other three in the savanna woodland ecosystem of southern DRC (Fig. 1). Even the single matched rainforest site is located over 600 km from the PA, at a much greater travel distance (multiple weeks by boat, the dominant means of transport in this region of DRC) from Kinshasa and its 17 million people.

Even more concerning is that the chosen SCs for site DRC 934 all have 0% 'Logging concession cover', whereas the project has 100% 'Logging Concession Cover (West et al. Table A17). This variable was matched for, but is not a strict condition, so SCs were allowed if they fitted other variables better even if they failed entirely to match this condition. However, forests located in or out of a logging concession likely have a totally different risk profile: the REDD project activity here was to prevent a logging concession becoming active.

This poor quality of matching is systematic: every PA in Peru that showed more deforestation than its SC was matched to an SC with less timber concession coverage; every Colombian PA that showed more deforestation than its SC was matched to an SC with a higher level of pre-existing legal protection; and over 80% of PAs were matched to an SC that had lower levels of deforestation in a 10 km buffer zone around it. Given these biases, a lack of significant difference in deforestation rates between PAs and SCs after project implementation is not surprising, and indeed could even itself indicate project success.

*Unreliable Validation Tests Used*

In their Figure S6, West et al. show a "validation" in which they divide the pre-treatment period into two parts (training and validation periods). If the method is reliable, then the SC and the PA should show parallel deforestation trends in the validation period, but this requirement appears to be violated. The authors claim that 26 out of 30 PAs pass the validation by setting a threshold in terms of absolute deforestation by the final year, equal to a deviation of less than 0.5% of the total project area. Replicating this same test, we find that 24 of the PAs pass using normal SCM, and 25 pass using an Augmented synthetic control method (ASCM) proposed by Ben-Michael et al. *(13)*.

However, this "validation" cannot be considered a rigorous test. The absolute size of error allowed is very large, especially for PAs with low deforestation rates (10 of the 32 PAs had pre-project total deforestation less than 0.5%, meaning a deviation of over 100% is allowed in this test).

The credibility of an SC estimator depends on its ability to track the trajectory of the outcome variable (in this case deforestation *d*) for an extended period in time *t (14)*, so a better numerical test is given by:

$$\frac{max(|d_{PA}(t)-d_{SC}(t)|)}{d_{PA}(t_{final})} < 0.2 \qquad \text{(Equation 3)}$$

In addition, following best practice *(9)* we applied a second test:

$$\frac{RMSPE(validation)}{RMSPE(training)} < 5 \qquad \text{(Equation 4)}$$

These equations and the derivation of values 0.2 and 5 as test statistics follow normal practice in the SC literature, see Supplementary Information for more detail.

We find that only 7 out of 30 PAs pass the requirement in Equation 3, and that these results are also more consistent with the graphical trends: only PAs that exhibit a parallel trend with the SC pass the test (Fig S1). For example, Project 1900 (Fig 2a) fails our test, despite passing that in West et al. We find that only 14 PAs pass the requirement of Equation 4, and suggest that if the authors truly wish to apply an "abundance of caution" *(1)*, then they should apply both tests, which 24 out of 30 PAs would fail. Our results show the same pattern when using ASCM: 6 PAs pass both validation tests, 4 of which also passed the tests using SCM (see supplementary materials).

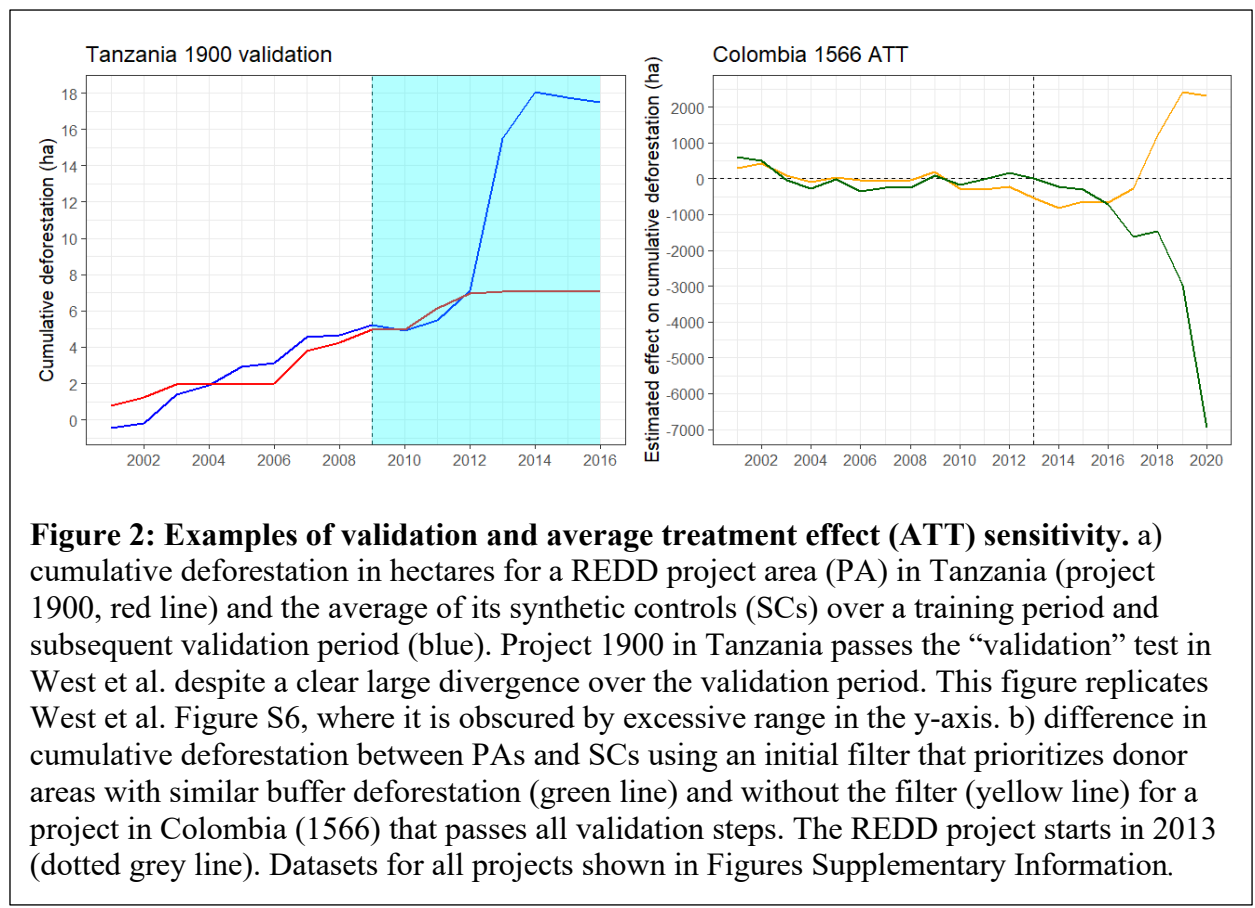

**Figure 2: Examples of validation and average treatment effect (ATT) sensitivity.** a) cumulative deforestation in hectares for a REDD project area (PA) in Tanzania (project 1900, red line) and the average of its synthetic controls (SCs) over a training period and subsequent validation period (blue). Project 1900 in Tanzania passes the "validation" test in West et al. despite a clear large divergence over the validation period. This figure replicates West et al. Figure S6, where it is obscured by excessive range in the y-axis. b) difference in cumulative deforestation between PAs and SCs using an initial filter that prioritizes donor areas with similar buffer deforestation (green line) and without the filter (yellow line) for a project in Colombia (1566) that passes all validation steps. The REDD project starts in 2013 (dotted grey line). Datasets for all projects shown in Figures Supplementary Information.

Given this failure rate (and arguably even their stated failure rate of over 10% of PAs), it should have been concluded that their SC model was not reliable and required adjustments to its input datasets, rather than continuing with the analysis. It is likely that improving the number of variables known to be important for deforestation, using a more reliable remote sensing datasets, and putting in hard controls related to biomes or land use types available for SC matches, would have improved these results.

*Sensitivity of Results*

Due to the flaws in the SCM approach discussed above, the results presented by West et al. are not robust to small changes in the algorithm used to generate SCs. This robustness testing is a common feature of SCM approaches, with reference 9 for example varying their models in many ways to see how the results change. To demonstrate this, we replicated their method with and without the application of a filter they used that prioritizes donor areas with similar rates of buffer deforestation. If the method were robust, we would expect the filter to have some effect, but for the overall results to be similar without it given that buffer deforestation remains one of the covariates.

In fact, removing the filtering step leads to reversal of outcomes for 7 of the PAs. For example, Colombia 1566 has 7,000 ha less deforestation than its SC when the filter is used, but 2,500 ha more deforestation than its SC when the filter is removed. On average, these numbers (which West et al. carry forward into their calculation of carbon credits), differ by over 200%. Such discrepancies clearly point to a lack of robustness that should have been addressed through improving the models, before drawing any conclusions about the effectiveness of the REDD projects using these methods.

This instability in results is seen to some degree in West et al's analysis. Their reported effectiveness for the same projects, using the same datasets, varies widely: for example overall effectiveness of all African projects covered varies from 0% (project-level SC analysis) to 16% (regional GSC) to 50% (regional GSC using cardinality matching by year 5, West et al. Table S12).

**Calculation Errors**

The authors make two further errors. Firstly, they unfairly compare their estimates of avoided deforestation to the pre-project estimates of carbon credits. This ignores the fact that carbon credits are actually issued *ex post* following an independently verified assessment of performance, and the number is normally lower than the original estimate due to reductions for leakage and residual deforestation.

In reality, the projects assessed by West et al., as stated in West et al. Table S4, issued 64,966,812 credits, but in their analysis West et al. compare to the 88,951,394 total in the *ex ante* baselines of Table S1 (note Table S4 records 0 issuances for project 2278, but we added 3,002,760 issued credits as now stated for the project on the Verra registry). As a result, West et al. understate the proportion of carbon credits that are 'real' according to their own analysis by 37%.

Secondly, when converting from avoided deforestation to carbon offsets (West et al. Table S1), the wrong numbers for "project baseline deforestation" are used. In all cases the deforestation between the year 2000 and the start of the project has been added to the baseline deforestation that occurred after the project started. This has the effect of inflating the baseline deforestation by a mean of 18.0% (range 0 - 91% with bigger disparities for more recent projects), and means that the average project was given an "expected carbon offset per hectare of avoided deforestation" value of 207 Mg $CO_2$ $ha^{-1}$, compared to a true value of 268 Mg $CO_2$ $ha^{-1}$ (Table S1). The exact magnitude of the error this causes varies by the set of projects West et al. consider to be successful in their different analyses, but we estimate it understates the carbon benefits of avoided deforestation by between 18 and 35%.

These two errors are independent, and should be combined to understand the overall impact of these calculation errors. If this is done with West et al.'s individual project based analysis, the 'proportion of credits that deliver real carbon benefits' should be increased by 61.6% (Table S2), purely due to calculation missteps. While this appears to have only a small impact on West et al.'s estimate of the overall effectiveness of the REDD projects in their analysis (an increase in effectiveness from 6.1% to 9.8% is still far from 100%), when combined with uncertainties resulting from the input satellite data and poorly selected comparison sites, all of which likely decreased the observed estimated impact of a truly impactful project, they are significant.

**Sample size**

West et al. extend their conclusions from the 24 REDD projects they study to all REDD projects. But there are 103 registered Verra REDD projects *(4)*, so they cover only 23%. Further these 24 are only 6% of the 377 REDD projects identified in the study by Atmadja et al. *(5)*, that West et al. cite. Worse still, their sample is not random, as projects were only included if they come from certain countries pre-selected by West et al. Standard statistical theory would suggest great caution in drawing conclusions about the whole set of REDD projects from this sample, instead limiting their conclusions to projects in the six countries covered.

**Conclusion**

West et al. conclude that "only a minority of projects achieved statistically significant reductions in comparison with *ex post* counterfactuals", but we have shown that this result is highly uncertain given the lack of sensitivity of their approach given their input satellite data, use of inappropriate synthetic controls resulting in poor synergies between control and project sites and sensitivity of their results to small changes in methodology, and calculation errors. As such, we believe their paper should be retracted or heavily revised.

It is certainly possible that some REDD projects have issued more credits than, in retrospect, was warranted: but that is not possible to prove using West et al.'s approach. We note that all these project areas have low rates of deforestation (average rate of 0.21%/yr according to West et al. figures, compared to an average of 0.47%/yr for forests in the six countries covered by West et al. *(7)*), despite often being located in higher risk areas.

We call for future studies on REDD project effectiveness that use locally tuned forest change data with known accuracies or point-based sampling approaches to quantify deforestation. Furthermore, analytical approaches must always lead to meaningful comparisons between forests of the same ecological type and legal status, and should consistently pass rigorous validation checks before conclusions are drawn from them. We suggest pixel based matching methods (e.g. *15, 16*) may be more appropriate given the volatile nature of deforestation as an outcome variable. Finally, we suggest that broad conclusions about avoided deforestation projects as a whole should not only satisfy these requirements, but also require a formal hypothesis testing framework and feature a representative sample of projects.

**Notes**

1. Using spatially proximate reference regions can introduce another problem, that of spatial autocorrelation, but this problem has been known for many years, and techniques have been developed in spatial statistics to nullify its effect. Further, unaccounted for spatial autocorrelation will in the case of such post hoc analyses of the effectiveness of REDD projects act conservatively: it will decrease the likelihood that the null hypothesis that PAs and control areas are the same is rejected.

**Acknowledgments:** We acknowledge responses from Thales West received in answer to emailed questions, which were helpful in preparing this study. David Coomes also provided useful feedback.

**Funding:** The authors' time spent preparing this study was funded by their institutions.

**Author contributions:**

    Conceptualization: ETAM

    Methodology: ETAM; RC; EN

    Investigation: ETAM; RC; EN; HC

    Visualization: HC

    Funding acquisition:

    Project administration: ETAM

    Supervision:

    Writing – original draft: ETAM; HC; EN; RC

    Writing – review & editing: All.

**Competing interests:** ETAM, HC, PNQ, TB and MC are employed by Space Intelligence, a company providing data for nature-based project identification, screening, and monitoring. EN is an independent (non-remunerated) expert member of the Verified Carbon Standard (VCS) Advisory Group at Verra. These affiliations demonstrate the authors' knowledge and passion for the topic, but did not influence the scientific rigor of their analysis.

**Data and materials availability:** All data used to replicate the results of West et al. is available at the following repository: https://doi.org/10.34894/IQC9LM (Accessed 01/11/2023)

# Supplementary Materials for

# Serious errors impair an assessment of forest carbon projects: A rebuttal of West et al. (2023)

**Authors:** Edward T. A. Mitchard[1,2]*, Harry Carstairs[1], Riccardo Cosenza[3],
Sassan S. Saatchi[4,5,6], Jason Funk[7], Paula Nieto Quintano[1], Thom Brade[1], Iain M. McNicol[1,2],
Patrick Meir[2], Murray B. Collins[1], Eric Nowak[3,8]

Corresponding author: ed@space-intelligence.com

**The PDF file includes:**

Materials and Methods
Tables S1 to S6
Figs. S1 to S6

## Materials and Methods

Validity of Synthetic Control approach

We conducted an empirical evaluation of the synthetic control method (SCM) employed by West et al. *(1)*. Our approach and the methodology are as close as possible, given the example scripts provided by the authors did not specify the years used to create training and validation periods or the exact cutoffs used for filtering potential donor areas by buffer deforestation. The analysis was conducted with the Synth and augsynth packages available for R software.

We replicate the results in West et al. Figure S6 (our Figure S1) and Table S3 (Table S3) using the dataset made available by West et al. in each REDD+ project using both the SCM (that they adopt) and the ASCM. The simple SCM used in their analysis does not allow negative weights while ASCM relaxes this constraint allowing for negativity and extrapolation. ASCM is particularly useful for improving the covariate balance between treated and control groups and often leads to a better pre-treatment fit *(13)*. Weights are chosen to balance pre-treatment outcomes and other covariates as in the SCM.

The first step when applying a SCM is to test the validity of the methodology. The approach consists of dividing the pre-treatment period in two parts (training and validation period) using a random year. This step is shown in fig. S3. We then apply three numerical tests to determine the quality of the fits in the validation period between project areas (PAs) and synthetic controls (SCs). In table S3 we replicate the same test used by West et al. We calculated the difference between project and synthetic control deforestation in the final year of the validation period and then we compared this number with the total project area. This is done for both the ASCM and the SCM.

In table S4 we applied a variant of the validation test proposed by Abadie et al. *(10)*. This test is based on the *ratio between the root mean squared prediction error (RMSPE) for the validation period and for the training period*. The smaller the value, the better is the validity of the SC. Based on experimentation and following guidance in the literature, we set 5 as a limit value for this ratio; if the *post RMSPE / Pre RMSPE ratio* is bigger than 5 the project fails the validation. The benefit of using this methodology is that it is not biased by the quality of the fit in the training period.
In addition, we proposed a second more reasonable numerical test using the *ratio between the maximum of the absolute value differences in the validation period and total deforestation*. This approach not only allows us to capture the divergence between project and SCs across the entire validation period, but also eliminates the bias of preference of low deforestation areas. Empirically we choose 20% (of *total deforestation in the last year of the validation period*) as a threshold to pass the validation test, as this seemed to keep only projects where by eye the match appeared strong in the validation period. The results are shown in table S5. We continued to use ASCM as this algorithm led to better validation results and therefore should be more robust.

Addressing the second concern around the sensitivity of SC results, we ran the analysis both with and without the filter applied by West et al. This filter selectively includes only those donors with deforestation patterns analogous to the project sites (ideally 10 percent deviation, but increasing this range to ensure an SC with a good match in the pre-implementation period is obtained). We used different percentage deviations (20 or 30 percent) for each country based on the ability of the SC to find enough controls for the analysis.

Figure S2 shows the cumulative deforestation of the PAs relative to the SCs created with and without the filter. We replicate West et al.'s figures S7 and S8 in our figures S3 and S4 (without filtering) and figures S5 and S6 (with filtering). Finally, in Table S6 we show the average treatment effects on the treated (ATT) for each project when the filter is applied and when it is not, which we defined as the mean difference between the deforestation in the PS and the SC over the implementation period.

**Table S1: Corrected carbon credits generated estimated using synthetic control (SC) results.**
Replication of West et al. table S5. The carbon offsets generated – according to their SC results – are calculated assuming a fixed value of carbon per hectare of avoided deforestation. This value is underestimated due to their incorrect addition of pre-implementation deforestation to the project baselines. Overall, this increases the number of credits generated by 18%.

| Project | Avoided Deforestation from West et al. SC results (ha) | Project Baseline Deforestation (ha) | | Carbon Credits Expected (Mg $CO_2$) | Carbon offsets per hectare (Mg $CO_2$ ha$^{-1}$) | | Carbon offsets from SC results (Mg $CO_2$) | |
|---|---|---|---|---|---|---|---|---|
| | | West et al. | Correct | | West et al. | Correct baseline | West et al. | Correct baseline |
| Peru 1882 | 0 | 1,268 | 1,016 | 356,960 | 282 | 351 | 0 | 0 |
| Peru 2278 | 0 | 13,299 | 12,581 | 5,891,253 | 443 | 468 | 0 | 0 |
| Peru 1067 | 4,727 | 13,581 | 11,893 | 4,817,471 | 355 | 405 | 1,676,768 | 1,914,755 |
| Peru 958 | 2,223 | 2,924 | 1,525 | 630,937 | 216 | 414 | 479,676 | 919,720 |
| Peru 944 | 3,457 | 23,685 | 20,842 | 5,151,165 | 217 | 247 | 751,850 | 854,408 |
| Peru 844 | 0 | 125,561* | 21,983 | 12,475,134 | 99 | 567 | 0 | 0 |
| Colombia 1400 | 0 | 12,425 | 12,157 | 1,657,098 | 133 | 136 | 0 | 0 |
| Colombia 1566 | 0 | 103,908 | 94,345 | 31,325,923 | 301 | 332 | 0 | 0 |
| Colombia 1396 | 2,992 | 8,076 | 7,417 | 1,489,786 | 184 | 201 | 551,937 | 600,976 |
| Colombia 1395 | 1,440 | 16,835 | 15,944 | 2,791,723 | 166 | 175 | 238,793 | 252,138 |
| Colombia 1392 | 117 | 10,722 | 10,633 | 1,455,141 | 136 | 137 | 15,879 | 16,012 |
| Cambodia 904 | 0 | 21,252 | 20,142 | 1,626,420 | 77 | 81 | 0 | 0 |
| Cambodia 1650 | 4,110 | 30,446 | 28,304 | 12,432,277 | 408 | 439 | 1,678,272 | 1,805,280 |
| DRC 1359 | 0 | 11,949 | 10,030 | 4,735,361 | 396 | 472 | 0 | 0 |
| Tanzania 1325 | 0 | 10,578 | 7,122 | 359,834 | 34 | 51 | 0 | 0 |
| Tanzania 1900 | 0 | 11,407 | 11,400 | 348,019 | 31 | 31 | 0 | 0 |
| Tanzania 1897 | 0 | 35,472 | 25,301 | 1,406,892 | 40 | 56 | 0 | 0 |
| **Sum** | | | | | | | **5,393,174** | **6,363,289** |

*The large disparity in Peru 844 is due to the wrong number for baseline deforestation being taken from the project documents.

**Table S2: Corrected proportion of carbon credits linked to reduced deforestation using SC results.**

Replication of the result in *(1)* that 6.1% of carbon offsets are linked to reduced deforestation. Here, we make two main corrections: firstly we use the correct values for the number of offsets generated per hectare (Table S1); and secondly we compare to credits actually *issued* (using values from West et al. Table S4) instead of *expected* credits (as used in West et al. Table S5). Combined, this increases the proportion of "real" carbon credits according to their SC analysis by 62%.

| Project | Carbon Offsets from project baseline (Mg CO$_2$) | | Carbon offsets from SC method (Mg CO$_2$) | | Percentage of credits "real" | |
|---|---|---|---|---|---|---|
| | Expected | Issued | West et al. | Correct baseline | West et al. | Correct baseline |
| Peru 1882 | 356,960 | 171,673 | 0 | 0 | 0% | 0% |
| Peru 2278 | 5,891,253 | 3,002,760* | 0 | 0 | 0% | 0% |
| Peru 1067 | 4,817,471 | 3,678,270 | 1,676,768 | 1,914,755 | 35% | 52% |
| Peru 958 | 630,937 | 566,843 | 479,676 | 919,720 | 76% | >100% |
| Peru 944 | 5,151,165 | 5,282,313 | 751,850 | 854,408 | 15% | 16% |
| Peru 844 | 12,475,134 | 9,658,069 | 0 | 0 | 0% | 0% |
| Colombia 1400 | 1,657,098 | 544,278 | 0 | 0 | 0% | 0% |
| Colombia 1566 | 31,325,923 | 22,274,745 | 0 | 0 | 0% | 0% |
| Colombia 1396 | 1,489,786 | 567,286 | 551,937 | 600,976 | 37% | >100% |
| Colombia 1395 | 2,791,723 | 1,620,202 | 238,793 | 252,138 | 9% | 16% |
| Colombia 1392 | 1,455,141 | 477,432 | 15,879 | 16,012 | 1% | 3% |
| Cambodia 904 | 1,626,420 | 48,000 | 0 | 0 | 0% | 0% |
| Cambodia 1650 | 12,432,277 | 14,568,314 | 1,677,170 | 1,805,280 | 13% | 12% |
| DRC 1359 | 4,735,361 | 1,620,202 | 0 | 0 | 0% | 0% |
| Tanzania 1325 | 359,834 | 10,000 | 0 | 0 | 0% | 0% |
| Tanzania 1900 | 348,019 | 150,425 | 0 | 0 | 0% | 0% |
| Tanzania 1897 | 1,406,892 | 726,000 | 0 | 0 | 0% | 0% |
| **TOTAL** | **88,951,394** | **64,966,812** | **5,392,073** | **6,363,289** | **6.06%** | **9.79%** |

*note West et al Table S4 records 0 issuances for project 2278, but we added 3,002,760 issued credits as now stated for the project on the Verra registry

**Table S3: Replication of West et al. Table S3.**

Synthetic control "validation": the difference between project and synthetic control deforestation in the final year of the validation period. The Difference in % is the % of the project area. Projects with a difference between project and synthetic control deforestation > 0.5% of the project area are assumed to have failed validation and are marked with an asterisk.

| Country | Project ID | End of validation period | ASCM | | SCM | |
|---|---|---|---|---|---|---|
| | | | Difference (ha) | Difference (%) | Difference (ha) | Difference (%) |
| Cambodia | 904 | 2007 | -272.5 | -0.4 | -497.6 | *-0.8 |
| Cambodia | 1650 | 2010 | -3,122.7 | *-1.6 | -702.9 | -0.4 |
| Colombia | 856 | 2011 | 20.8 | 0.2 | 54.4 | 0.4 |
| Colombia | 1389 | 2013 | -1,994.8 | *-2.9 | -1,420.7 | *-2.1 |
| Colombia | 1390 | 2014 | 127.7 | 0.1 | 348.0 | 0.3 |
| Colombia | 1391 | 2013 | -102.5 | -0.2 | 67.9 | 0.1 |
| Colombia | 1392 | 2013 | -26.9 | 0.0 | -60.5 | -0.1 |
| Colombia | 1395 | 2013 | 257.9 | 0.3 | 61.0 | 0.1 |
| Colombia | 1396 | 2014 | -36.9 | -0.1 | 67.1 | 0.1 |
| Colombia | 1400 | 2013 | -34.1 | -0.1 | 195.8 | 0.3 |
| Colombia | 1566 | 2013 | 553.8 | 0.0 | -842.7 | 0.0 |
| DRC | 934 | 2011 | -2,052.4 | *-0.7 | 56.3 | 0.0 |
| DRC | 1359 | 2009 | -826.6 | -0.4 | -2,946.0 | *-1.6 |
| Peru | 844 | 2009 | -10.0 | 0.0 | 0.3 | 0.0 |
| Peru | 944 | 2009 | 275.3 | 0.2 | 368.6 | 0.2 |
| Peru | 958 | 2011 | 212.8 | 0.1 | -7.5 | 0.0 |
| Peru | 985 | 2009 | -1,008.8 | -0.1 | -26,682.6 | *-2.0 |
| Peru | 1067 | 2011 | -829.9 | -0.1 | -1,254.1 | -0.2 |
| Peru | 1182 | 2013 | -48.2 | -0.1 | -74.2 | -0.1 |
| Peru | 2278 | 2018 | -863.8 | -0.5 | -611.2 | -0.3 |
| Peru | 1360-1 | 2010 | -1,198.0 | *-1.3 | -6,706.3 | *-7.5 |
| Peru | 1360-2 | 2010 | -27.3 | -0.1 | -3.1 | 0.0 |
| Peru | 1360-3 | 2010 | 5.8 | 0.0 | 11.5 | 0.1 |
| Tanzania | 1325 | 2011 | 351.5 | 0.5 | 480.4 | 0.7* |
| Tanzania | 1897 | 2017 | 4,685.5 | *2.3 | 3,396.6 | *1.7 |
| Tanzania | 1900 | 2016 | -10.4 | 0.0 | -260.6 | -0.2 |
| Zambia | 1202 | 2009 | -53.4 | -0.1 | -24.4 | -0.1 |
| Zambia | 1775-1 | 2015 | -3,147.8 | -0.5 | -2,446.0 | -0.4 |
| Zambia | 1775-2 | 2015 | -310.2 | -0.2 | 290.5 | 0.2 |
| Zambia | 1775-3 | 2015 | -252.4 | -0.1 | -177.2 | -0.1 |

**Table S4: Post RMSPE / Pre RMSPE ratios.**

Validation test. ratio between the root mean squared prediction error (RMSPE) for the training period and for the validation period. Projects with a Post RMSPE / Pre RMSPE > 5 are assumed to have failed validation and are marked with an asterisk.

|  |  | ASCM | SCM |
|---|---|---|---|
| *Country* | *Project ID* | *Post RMSPE / Pre RMSPE* | *Post RMSPE / Pre RMSPE* |
| Cambodia | 904 | *11.0 | 3.8 |
| Cambodia | 1650 | *7.8 | 1.4 |
| Colombia | 856 | *5.5 | *9.3 |
| Colombia | 1389 | *203.4 | *7.4 |
| Colombia | 1390 | 5.0 | *44.0 |
| Colombia | 1391 | 4.6 | * 8.9 |
| Colombia | 1392 | *6.4 | *26.5 |
| Colombia | 1395 | *17.1 | *6.2 |
| Colombia | 1396 | 1.0 | 4.8 |
| Colombia | 1400 | 3.5 | *45.7 |
| Colombia | 1566 | 1.5 | 4.2 |
| DRC | 934 | *85.0 | 3.0 |
| DRC | 1359 | 4.3 | 3.2 |
| Peru | 844 | *12.3 | 1.0 |
| Peru | 944 | 3.0 | 4.4 |
| Peru | 958 | 2.9 | 0.9 |
| Peru | 985 | 4.7 | 2.5 |
| Peru | 1067 | 2.2 | 3.1 |
| Peru | 1182 | 4.3 | *6.0 |
| Peru | 2278 | *13.1 | *9.9 |
| Peru | 1360-1 | *18.3 | 2.8 |
| Peru | 1360-2 | 4.1 | 1.5 |
| Peru | 1360-3 | 4.8 | *6.0 |
| Tanzania | 1325 | *9.8 | *6.3 |
| Tanzania | 1897 | *42.1 | *25.6 |
| Tanzania | 1900 | *7.8 | *18.0 |
| Zambia | 1202 | *5.6 | 3.3 |
| Zambia | 1775-1 | *10.5 | *13.3 |
| Zambia | 1775-2 | *6.1 | *11.3 |
| Zambia | 1775-3 | *9.7 | *27.6 |

## Table S5: Synthetic control validation.

Synthetic control validation: the maximum of the absolute value differences between project and synthetic control deforestation in each year of the validation period. The Difference in % is the % of the project deforestation in the final year of the validation period. Projects with a % difference > 20% are assumed to have failed validation and are marked with an asterisk.

| | | | ASCM | | SCM | |
|---|---|---|---|---|---|---|
| *Country* | *Project ID* | *years* | *Difference (ha)* | *Difference (%)* | *Difference (ha)* | *Difference (%)* |
| Cambodia | 904 | 2004-2007 | 272.5 | *26.4 | 497.6 | *48.2 |
| Cambodia | 1650 | 2007-2010 | 3,122.7 | *145.8 | 702.9 | *32.8 |
| Colombia | 856 | 2006-2011 | 20.8 | 14.1 | 54.4 | *36.8 |
| Colombia | 1389 | 2008-2013 | 1,994.8 | *215.2 | 1,420.7 | *153.2 |
| Colombia | 1390 | 2008-2014 | 127.7 | *22.6 | 348.0 | *61.5 |
| Colombia | 1391 | 2008-2013 | 102.5 | *53.7 | 90.4 | *47.3 |
| Colombia | 1392 | 2007-2013 | 26.9 | *30.2 | 60.5 | *67.9 |
| Colombia | 1395 | 2008-2013 | 276.0 | *31.0 | 228.2 | *25.6 |
| Colombia | 1396 | 2008-2014 | 36.9 | 5.6 | 69.1 | 10.5 |
| Colombia | 1400 | 2007-2013 | 42.9 | 16.0 | 195.8 | *72.9 |
| Colombia | 1566 | 2007-2013 | 553.8 | 5.8 | 851.5 | 8.9 |
| DRC | 934 | 2006-2011 | 2,052.5 | *24.2 | 667.6 | 7.9 |
| DRC | 1359 | 2006-2009 | 826.6 | *43.1 | 2,946.0 | *153.5 |
| Peru | 844 | 2005-2009 | 10.0 | 16.6 | 5.6 | 9.3 |
| Peru | 944 | 2005-2009 | 326.6 | 11.5 | 507.9 | 17.9 |
| Peru | 958 | 2006-2011 | 212.8 | 12.2 | 109.5 | 6.3 |
| Peru | 985 | 2006-2009 | 1,723.0 | *77.5 | 26,682.6 | *1,199.6 |
| Peru | 1067 | 2007-2011 | 882.4 | *52.3 | 1,254.1 | *74.3 |
| Peru | 1182 | 2007-2013 | 48.2 | 19.1 | 74.2 | *29.4 |
| Peru | 2278 | 2011-2018 | 863.8 | *120.2 | 611.2 | *85.1 |
| Peru | 1360-1 | 2007-2010 | 1,198.0 | *328.4 | 6,706.3 | *1,838.3 |
| Peru | 1360-2 | 2007-2010 | 27.3 | *86.4 | 14.1 | *44.6 |
| Peru | 1360-3 | 2006-2010 | 22.2 | *30.1 | 21.2 | *28.8 |
| Tanzania | 1325 | 2006-2011 | 468.0 | 13.5 | 627.5 | 18.2 |
| Tanzania | 1897 | 2010-2017 | 4,685.5 | *46.1 | 3,396.6 | *33.4 |
| Tanzania | 1900 | 2009-2016 | 11.0 | *155.4 | 260.6 | *3,691.6 |
| Zambia | 1202 | 2006-2009 | 53.4 | *131.4 | 36.8 | *90.6 |
| Zambia | 1775-1 | 2009-2015 | 3,147.8 | *395.8 | 2,329.0 | *292.8 |
| Zambia | 1775-2 | 2009-2015 | 310.2 | *67.9 | 290.5 | *63.6 |
| Zambia | 1775-3 | 2009-2015 | 252.4 | *159.9 | 177.2 | *112.2 |

**Table S6: Averages of the average treatment effects on treated (ATT).**
Differences between the averages ATT for each project when the filter is applied and when it is not. The ATT is calculated as the mean difference between the deforestation in the project area and the synthetic control deforestation over the implementation period.

| Country | Project ID | Without filter | With filter | Difference (%) |
| --- | --- | --- | --- | --- |
| Cambodia | 1650 | -9,375.14 | -17,464.24 | 46% |
| Cambodia | 904 | -1,385.98 | -5,709.93 | 76% |
| Colombia | 1392 | -49.20 | -15.37 | -220% |
| Colombia | 1566 | 504.60 | -2,035.37 | 125% |
| Colombia | 856 | -3.22 | 76.51 | 104% |
| Colombia | 1391 | -188.37 | -98.99 | -90% |
| Colombia | 1396 | -464.40 | -404.86 | -15% |
| Colombia | 1400 | -353.82 | -364.89 | 3% |
| Colombia | 1389 | -4,506.85 | -3,019.10 | -49% |
| Colombia | 1395 | -582.62 | -820.55 | 29% |
| Colombia | 1390 | -107.12 | 182.65 | 159% |
| DRC | 1359 | 837.89 | -3,169.86 | 126% |
| DRC | 934 | -1,261.76 | 379.42 | 433% |
| Peru | 1360-3 | -94.74 | -132.38 | 29% |
| Peru | 844 | -697.61 | -35.25 | -1879% |
| Peru | 1067 | -1,545.24 | -6432.88 | 76% |
| Peru | 944 | -2,541.10 | -2922.78 | 13% |
| Peru | 985 | -9,873.20 | -108971.30 | 91% |
| Peru | 958 | -450.49 | -854.08 | 47% |
| Peru | 1360-2 | -122.58 | -274.98 | 55% |
| Peru | 2,278 | -567.84 | -506.51 | -12% |
| Peru | 1360-1 | -28.20 | -73.12 | 61% |
| Peru | 1182 | -99.55 | -125.87 | 21% |
| Tanzania | 1325 | 1,604.99 | 47.46 | -3282% |
| Tanzania | 1897 | 1,794.07 | 1,985.91 | 10% |
| Tanzania | 1900 | -0.01 | 0.19 | 105% |
| Zambia | 1775-1 | -4,121.52 | -4,337.38 | 5% |
| Zambia | 1775-2 | -72.75 | 328.68 | 122% |
| Zambia | 1775-3 | -236.41 | -316.88 | 25% |
| Zambia | 1202 | -62.27 | -1,107.55 | 94% |

**Figure S1: Replication of Figure S6.**
"Validation" of the synthetic control method in each REDD+ project using an augmented synthetic control method (ASCM). Pre-REDD+ deforestation in "to-be" REDD+ project areas (red) versus synthetic controls (blue) Shaded blue areas represent the validation periods.

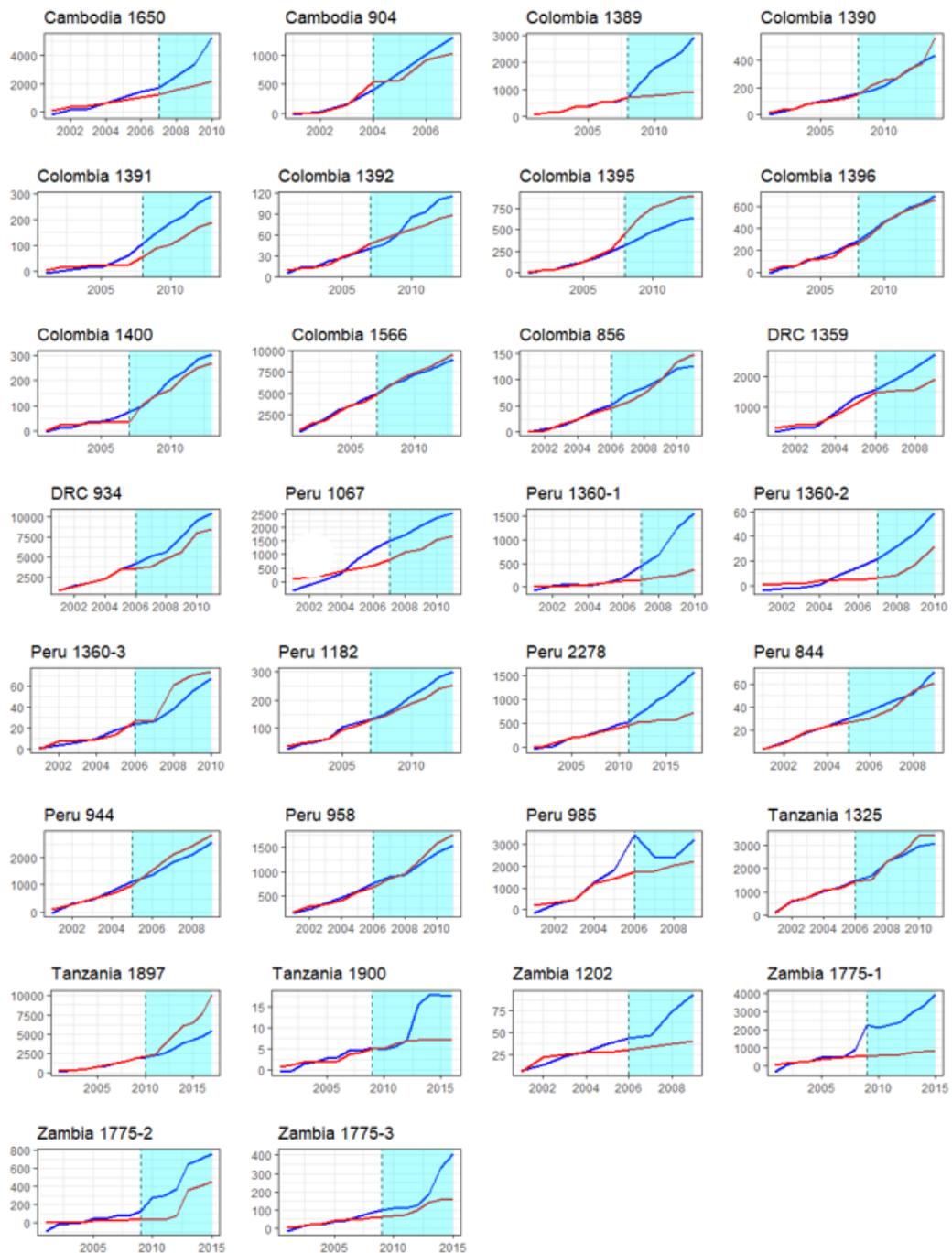

**Figure S2: Sensitivity of average treatment effects on treated (ATT).**
Differences between the averages ATT for each project when the filter is applied (green line) and when it is not (orange line). The ATT is calculated as the difference between the deforestation in the project area and the synthetic control deforestation.

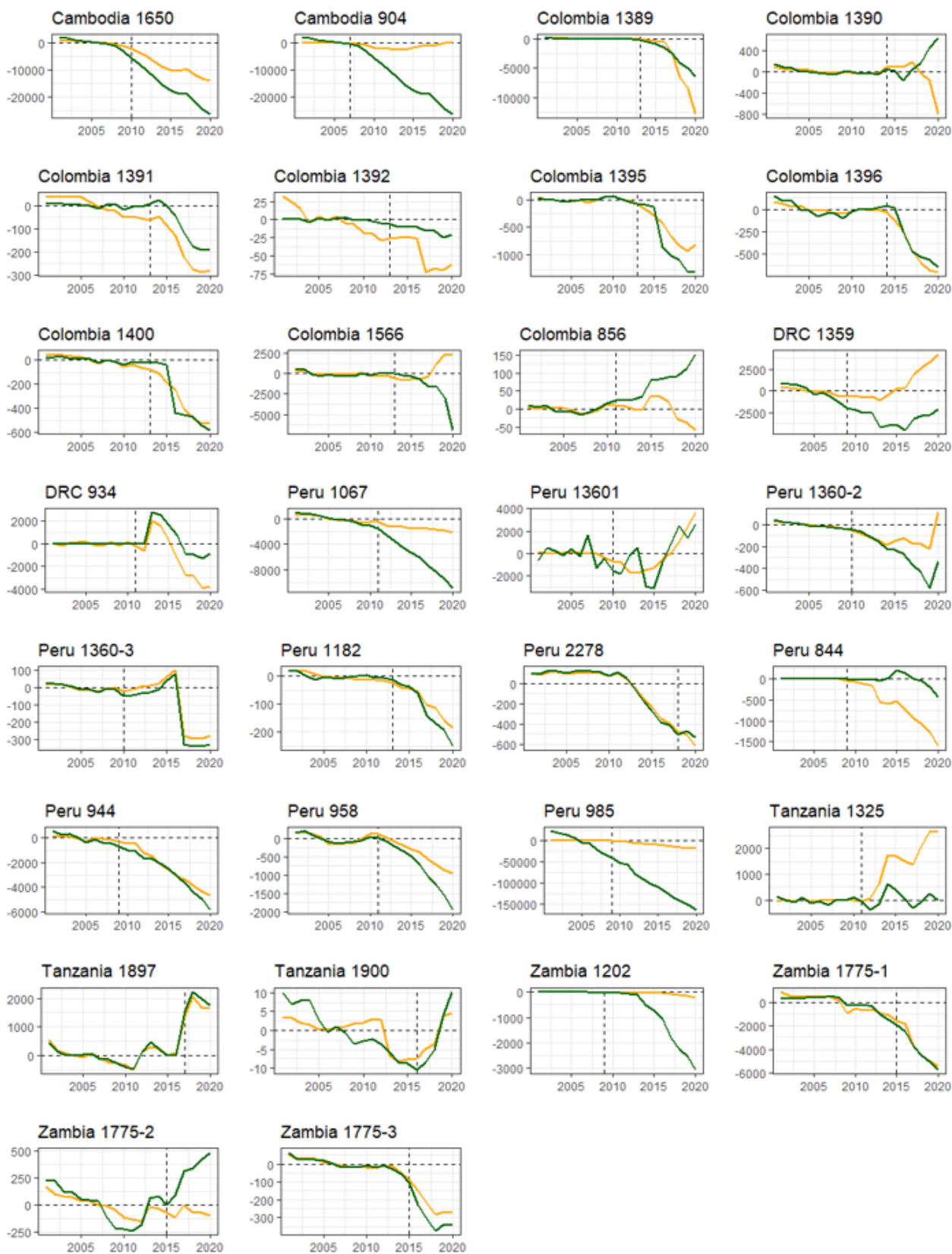

**Figure S3: Replication of Figure S7. (without filter)**
Cumulative post-2000 deforestation in REDD+ project areas (red) versus synthetic controls (blue). Dashed black lines indicate the project implementation year.

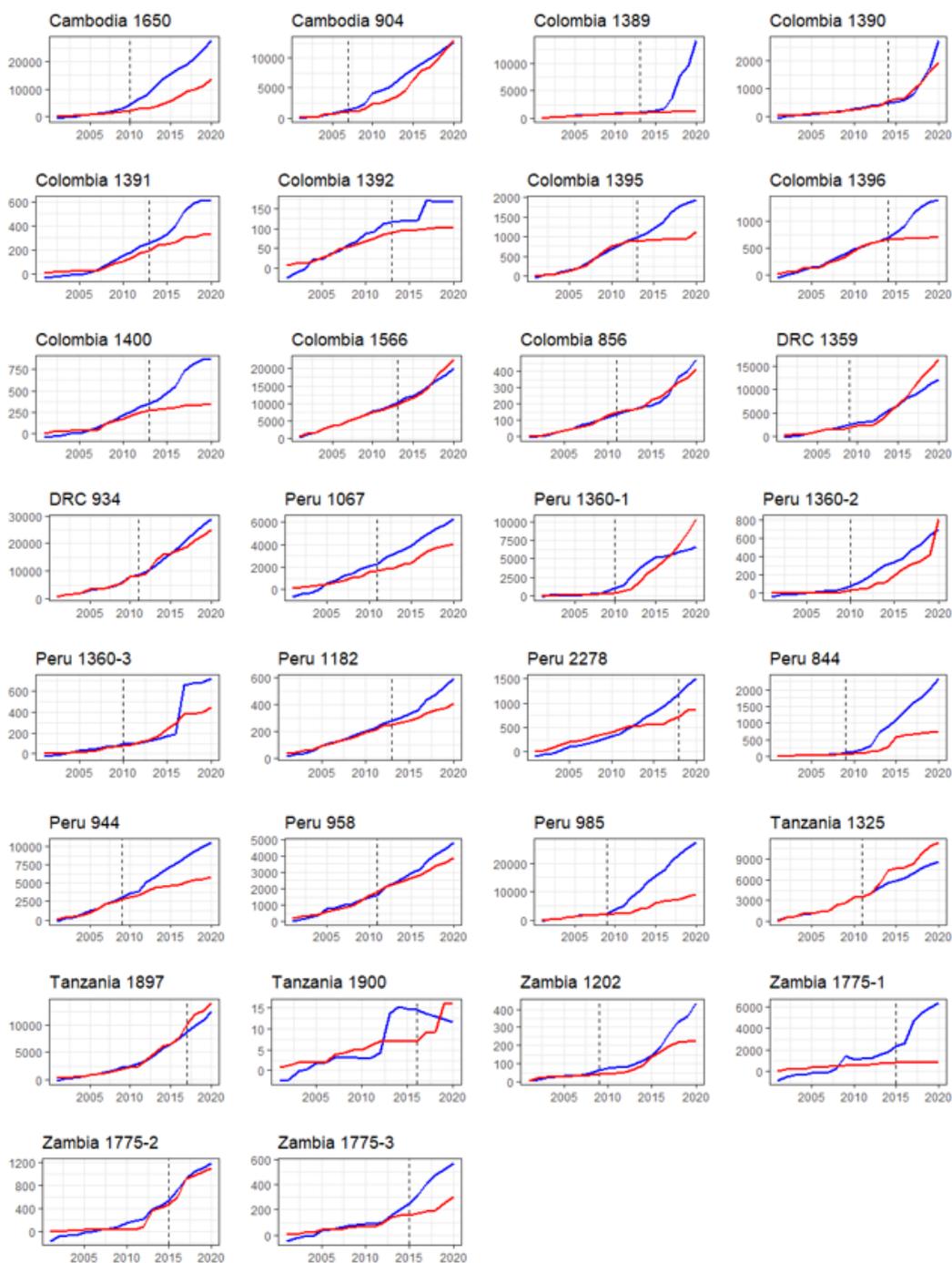

**Figure S4: Replication of Figure S8. (without filter)**
Cumulative deforestation in REDD+ project areas minus deforestation in their respective synthetic controls. Dashed black lines indicate the project implementation year. Grey shaded areas represent 95% confidence intervals, obtained using the jackknife+ method *(17)*.

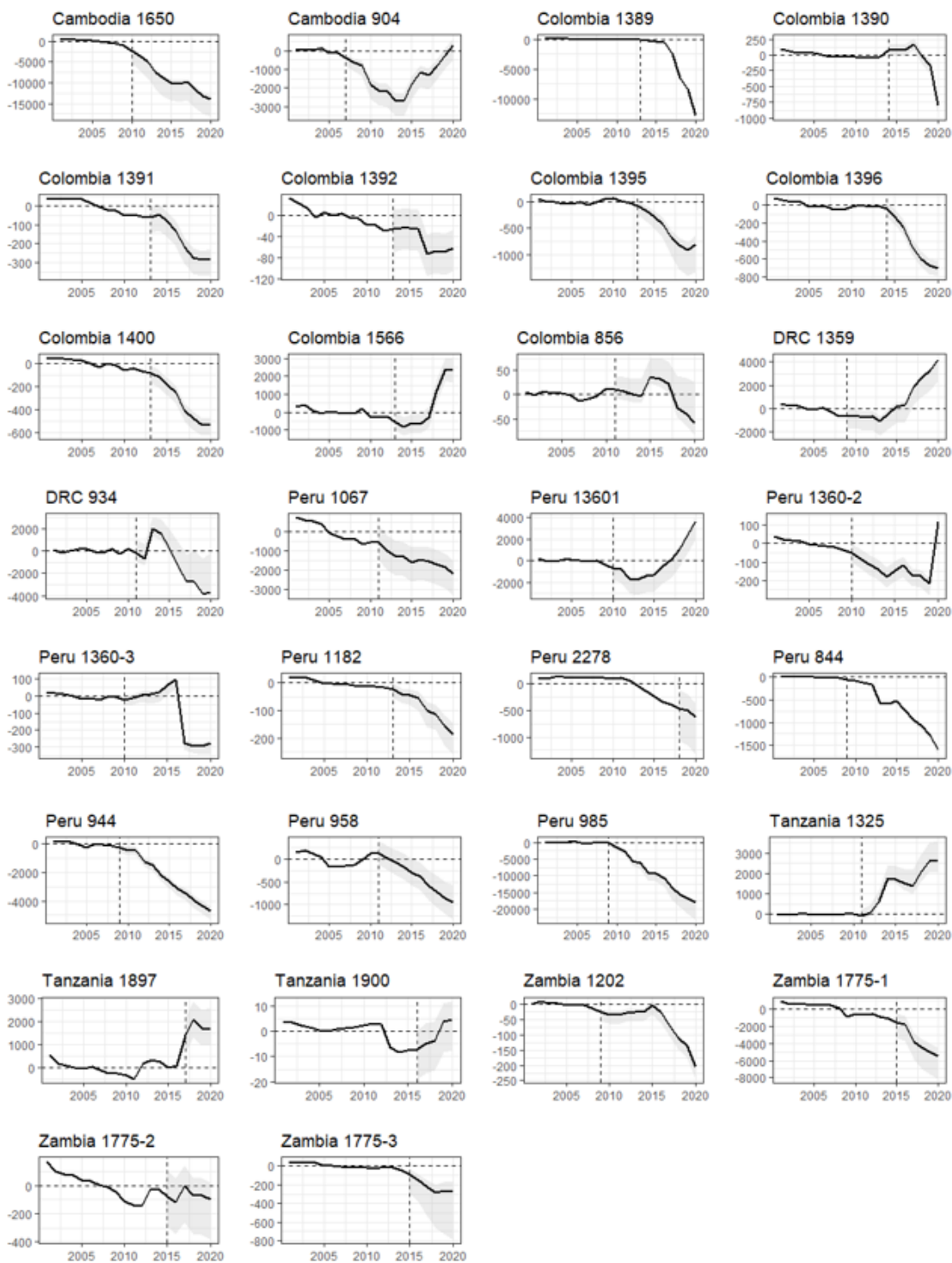

# Figure S5: Replication of Figure S7. (with filter)

Cumulative post-2000 deforestation in REDD+ project areas (red) versus synthetic controls (blue). Dashed black lines indicate the project implementation year.

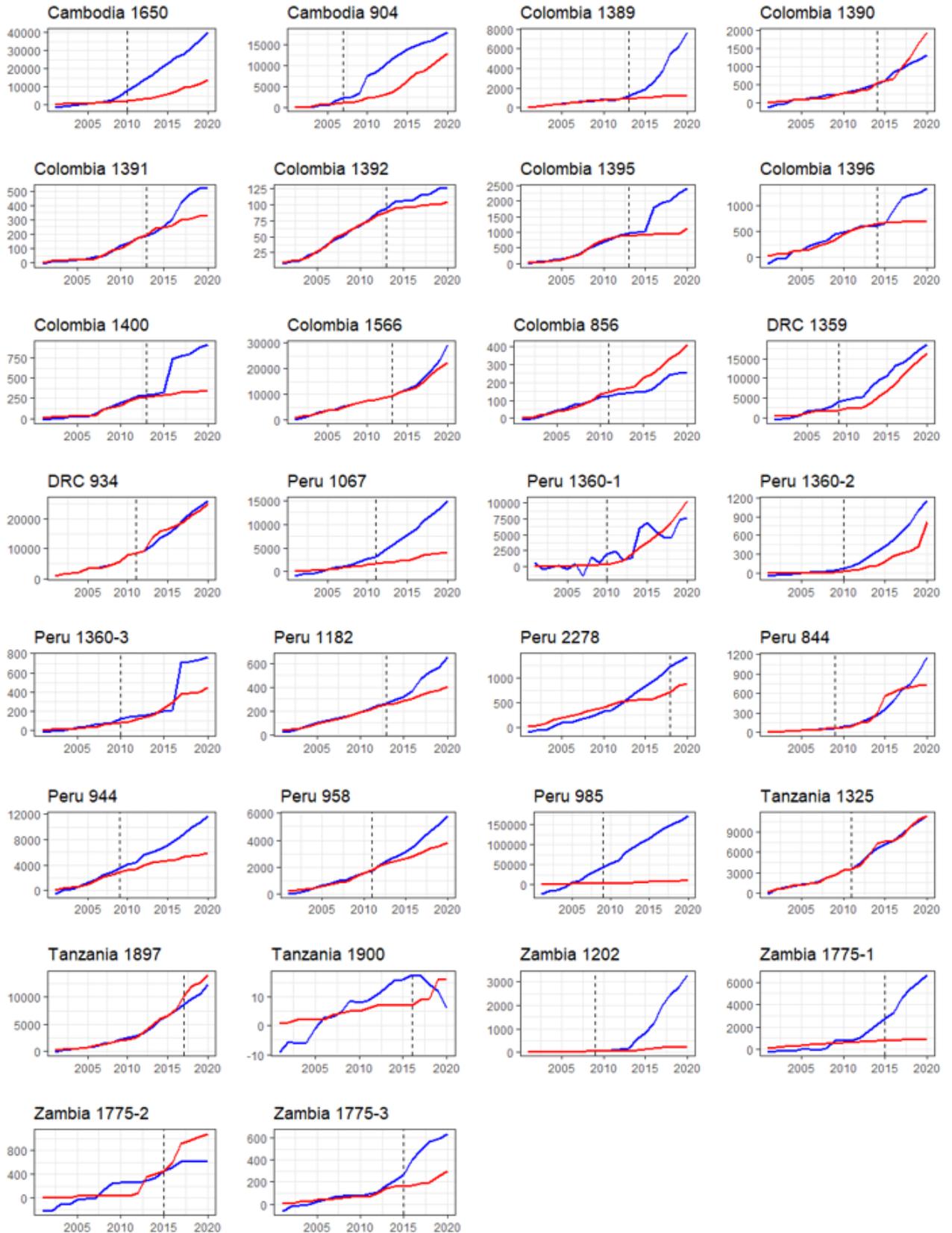

**Figure S6: Replication of Figure S8. (with filter)**
Cumulative deforestation in REDD+ project areas minus deforestation in their respective synthetic controls. Dashed black lines indicate the project implementation year. Grey shaded areas represent 95% confidence intervals, obtained using the jackknife+ method as in Barber et al. (2021)

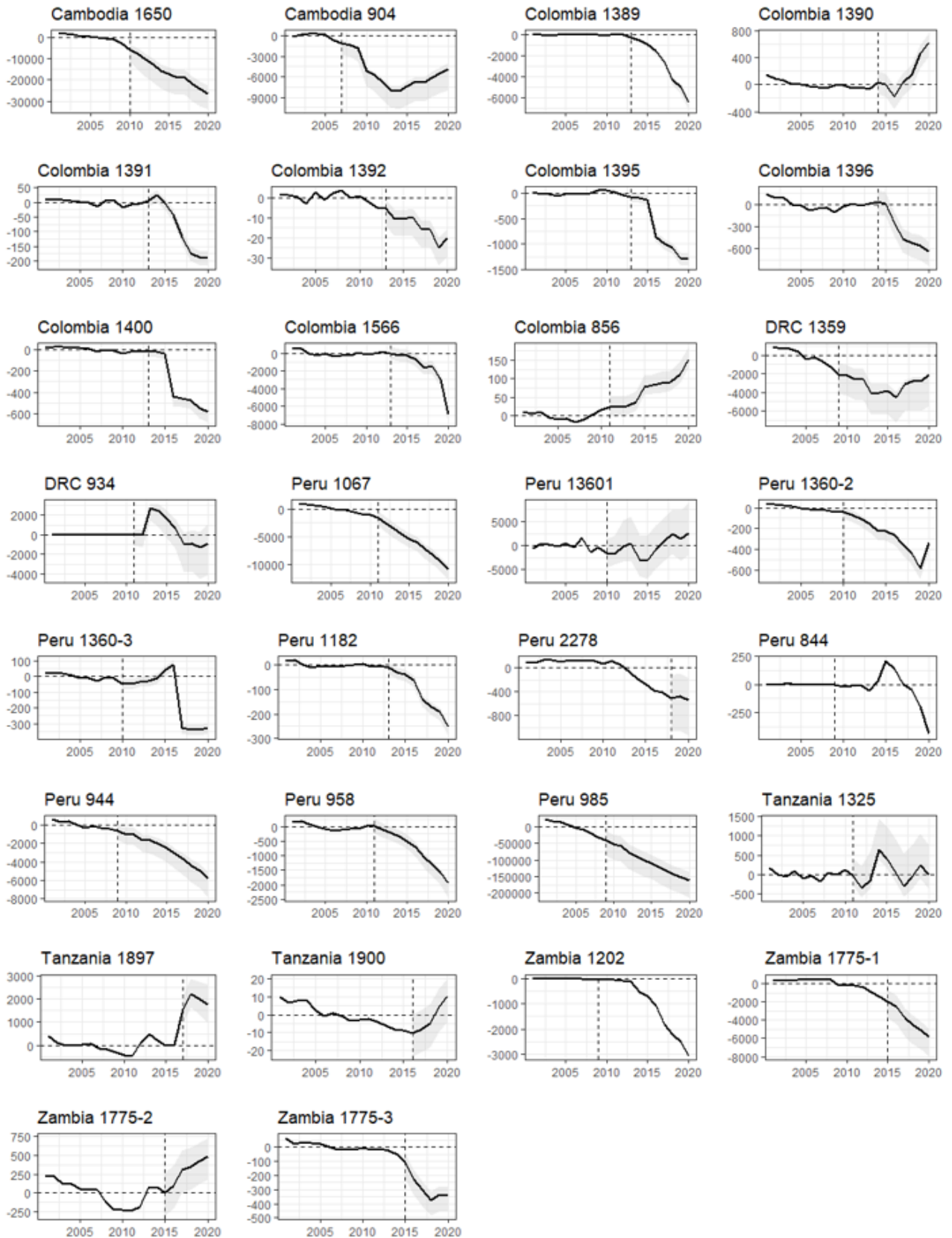